\documentclass[preprintnumbers,amsmath,amssymb,aps,pre]{revtex4}
\usepackage{mathptmx}      
\usepackage{nicefrac}         
\usepackage{appendix}      
\usepackage{comment}      
\usepackage{amsmath} 
\usepackage{graphicx} 
\begin{document}

\title{The role of noise and initial conditions in the asymptotic solution of a bounded confidence, continuous-opinion model}

\author{Adri\'an Carro, Ra\'ul Toral,Maxi San Miguel}

\affiliation{ Instituto de F\'isica Interdisciplinar y Sistemas Complejos IFISC (CSIC-UIB),
              Campus Universitat Illes Balears, E07122, Palma de Mallorca, Spain }

\begin{abstract}
We study a model for continuous-opinion dynamics under bounded confidence. In particular, we analyze the importance of the initial distribution of opinions in determining the asymptotic configuration. Thus, we sketch the structure of attractors of the dynamical system, by means of the numerical computation of the time evolution of the agents density. We show that, for a given bound of confidence, a consensus can be encouraged or prevented by certain initial conditions. Furthermore, a noisy perturbation is added to the system with the purpose of modeling the \textit{free will} of the agents. As a consequence, the importance of the initial condition is partially replaced by that of the statistical distribution of the noise. Nevertheless, we still find evidence of the influence of the initial state upon the final configuration for a short range of the bound of confidence parameter.

\end{abstract}
\maketitle



\section{Introduction}
\label{introduction}

The models of ``opinion dynamics'' focus on the processes of opinion formation within a society consisting of an ensemble of interacting individuals with diverse opinions. A wide variety of models, inspired by statistical mechanics and nonlinear physics, have been developed in order to deal with the different phenomena observed in real societies \cite{Castellano2009}: emergence of fads, minority opinion survival and spreading, collective decision making, emergence of extremism and so on. One of the main problems addressed by some of these models is whether the opinion formation processes within a society will eventually lead to the emergence of a consensus, with a vast majority of the agents adopting the same opinion, or to the fragmentation of its constituent individuals into different opinion groups. 

In opinion dynamics, each agent is characterized by an opinion from a certain space, which is a dynamical variable evolving in accordance with some rules. These behavioral laws codify a variety of internal and external factors governing the evolution of an agent's opinion, such as the social influence of its acquaintances, the social pressure of a group or the influence of advertising. Regarding the opinion variable, models can be broadly classified as \textit{discrete opinion models}, where opinions can only adopt a finite set of values, and \textit{continuous opinion models}, where opinions are real numbers in a finite interval and thus two interacting agents can always reach a compromise in an intermediate opinion.

Discrete models have traditionally been predominant in the physics literature, due to their correspondence with spin systems. They have been applied to analyze situations where individuals are confronted with a limited number of options, such as choosing among a few political parties in an election or between two languages in a language competition situation. On the contrary, continuous models are applied when a single issue is considered and opinions can vary continuously, for example, from ``completely against'' to ``in complete agreement''. Typical examples of continuous opinion issues are the degree of agreement regarding the legalization of drugs or abortion, or predictions about macroeconomic variables.

Two models of continuous opinion dynamics were introduced around 2000 and, ever since, have received much attention \cite{Lorenz2007}: the model of Hegselmann and Krause \cite{HK2002}, and that of Deffuant, Weisbuch \textit{et al.} \cite{Deffuant2000}, \cite{Weisbuch2002}. The former was first introduced in a mathematical context as a nonlinear version of older consensus models \cite{HK2000}, while the latter was developed in the context of a European Union project for the improvement of agri-environmental policies. Both models implement basically two mechanisms or rules for the evolution of the agent's opinion variables. On the one hand, there is a mechanism of social influence, by which two interacting agents tend to bring their opinions closer and, eventually, they reach a compromise in the midpoint opinion. On the other hand, both models take into account a mechanism of homophily, in particular a \textit{bounded confidence} rule, in the sense that agents do only interact if their opinion difference is less than a given threshold value. In other words, an agent will only take into account the opinions of other agents if they differ less than a bound of confidence $\epsilon$ from its own opinion, simply ignoring the rest of them.

The main difference between Hegselmann-Krause model and the Deffuant, Weisbuch \textit{et al.} model (from now on, DW \textit{et al.}) is the communication regime. In the case of DW \textit{et al.}, the communication is pairwise, i.e., agents meet in random pairwise encounters after which they compromise or not. On the contrary, in the Hegselmann-Krause model the communication takes place in groups, as each agent moves its own opinion to the average of the opinions of all the agents within a bound of confidence, including itself.

In the following, we will focus on the DW \textit{et al.} dynamics, which leads to final states where either a perfect consensus has been achieved, or the individuals split into a finite number of opinion clusters, depending on a parameter representing the confidence bound of the agents. In this paper, we study the influence that the initial distribution of opinions among the agents has upon the configuration of these final states. The fundamental question to answer is: can we, by imposing a given initial condition, force the system to arrive at a consensus, or, equivalently, prevent a consensus and force it to split in several opinion groups? Thus, we try to prove that the use of different initial distributions does not only have an effect in the average final opinion, but also in the fact that a consensus is found or not for a certain threshold level. To this end, we analyze, by means of numerical simulation, the effect of introducing a bias in the initial distribution towards the extremes or the center of the opinion space, with a parameter which allows for a continuous change between these two situations. In order to check the impact of the symmetry of the initial distribution, this analysis is performed both in a symmetric and in an asymmetric context. We find that, for a given value of the confidence bound parameter, the initial condition has a strong importance in determining the final state of the system, not only the position of the final opinion groups being affected, but also the total number of these clusters. Furthermore, we also analyze the influence of the initial conditions in the case of a DW \textit{et al.} model including an additional element of randomness in the form of a noise \cite{Raul2009,Raul2011}. This noisy term can be thought of as a \textit{free will}, since the agents are given the opportunity to change their opinion independently of their acquaintances. Concerning this noisy case, we find that, even if the noise hides most of the importance of the initial distribution, the latter has still some noticeable effects upon the final configuration.

The next section presents the DW \textit{et al.} model in greater depth, as well as a brief review of some results published in recent years regarding this model and its extensions to take into account the \textit{free will} of the agents by adding noisy perturbations. In section \ref{measures} we present the different initial conditions tested and the measures used to characterize these distributions and to quantify the results obtained. We present as well the general features of the simulations performed. Sections \ref{originalRes} and \ref{noisyRes} are devoted to the presentation of the results. In the first one we study the importance of the initial distribution of opinions for the stationary result of the original DW \textit{et al.} model, while in the second one we analyze the case in which a noise is added to the system. We draw, in section \ref{conclusions}, some general conclusions from the results of our study.



\section{The DW \textit{et al.} model and its noisy extension}
\label{models}

We will review, in a first subsection, the model developed in references \cite{Deffuant2000}, \cite{Weisbuch2002} in its original agent-based form, i.e., with its dynamics or behavioral rules defined for a finite population of $N$ agents. We will also present a redefinition of the system as a density-based model, as introduced by Ben-Naim, Krapivsky and Redner \cite{BenNaim2003} and by Lorenz \cite{Lorenz2005Pro}, where the dynamics is defined on the density of agents in the opinion space. We have used this second approach for the computation of all the results presented in this paper. Finally, we will describe in subsection \ref{noisyMod} the introduction of noise in the original model, in order to capture the \textit{free will} of the agents.


\subsection{The original model}
\label{originalMod}

In order to define the original agent-based version of the model, we begin by considering a group of $N$ agents, where we denote by $x_i (t)$ the real number representing the opinion that individual $i$ has at time step $t$ about a given topic. Without loss of generality we take $x_i (t) \in [0, 1]$. The behavioral rules defined for the agents basically state that, at each time step, a pair of individuals, say $i$ and $j$, is randomly chosen. Then, if their opinions are close enough, that is, if they satisfy $|x_i(t) - x_j(t)| < \epsilon$, being $\epsilon$ the bound of confidence, they are respectively adjusted as
\begin{eqnarray} 
  x_i(t+1) &= x_i(t) \, + \mu \left[ x_j(t) - x_i(t) \right],\nonumber\\[5pt]
  x_j(t+1) &= x_j(t) + \mu \left[ x_i(t) - x_j(t) \right],
\label{modelDef}
\end{eqnarray}
remaining unchanged otherwise. The iteration of this dynamical rule leads the system to a static final configuration which, depending on the parameters $\mu$ and $\epsilon$, can be a state of full consensus or one of fragmentation, where the individuals split in several opinion clusters of different sizes. This definition generates a mean field model, since all the agents have the same probability to be chosen as interaction partners.

The parameter $\mu$, which is restricted to the interval $(0, 0.5]$, can be thought of as the \textit{persuasibility} of the individuals, since it states how far an agent is willing to change its opinion. It basically fixes the speed of convergence and, because of this, it has some importance in determining the final number of clusters \cite{Laguna2004}, \cite{Porfiri2007}. In particular, for intermediate and large values of $\mu$, the speed of convergence will be so fast in the extremes of the opinion space that some small amount of agents will be left around that region with no possibility of communication with the rest of the individuals. These agents will therefore not be able to change and moderate their opinions, forming two minority clusters of extremists. On the contrary, for small values of $\mu$, i.e., for slow convergence speeds, all the agents will have the opportunity to meet and interact with others, being influenced by them, and thus slowly converging to the major clusters. Following most studies, and trying to avoid any further complexity, we will adopt from now on in this paper the value $\mu=0.5$, meaning a perfect compromise between interacting individuals.

The confidence parameter $\epsilon$ is restricted, for the opinion space introduced above, to the interval $\epsilon \in [0, 1]$. It is a measure of the receptiveness of the agents, that is, their willingness or readiness to interact with other individuals who have ideas different from theirs. It states how similar to me another individual must be so that I am willing to interact with him. In a typical realization of the agent-based dynamics starting from uniformly distributed random initial opinions, a consensus is obtained for large values of the confidence parameter, $\epsilon \geq 0.5$, while a fragmentation into opinion clusters separated by distances larger than $\epsilon$ can be observed for smaller values, whether these clusters are large or small.

The first density-based approach to the DW \textit{et al.} model, developed in \cite{BenNaim2003}, basically involved changing the scope from a finite number of agents to an idealized infinite number of individuals which are distributed in the opinion interval as defined by a density function. Therefore, the system is described in terms of a master equation for this density function $P(x,t)dx$, defined as the fraction of agents that have opinions in the range $[x, x + dx]$ at time $t$,
\begin{equation}
 \frac{\partial}{\partial t} P(t,x) = \int_0^1 dx_1 \int_{|x_1-x_2| \leq \epsilon} dx_2 \Bigg[ P(t,x_1) P(t,x_2) \bigg( \underbrace{2 \delta \left( x - \frac{x_1 + x_2}{2} \right)}_\text{Fraction joining state $x$} - \underbrace{(\delta (x - x_1) + \delta(x - x_2))}_\text{Fraction leaving state $x$} \bigg) \Bigg],
\label{originalME}
\end{equation}
where the persuasibility of the agents is taken as $\mu = 0.5$. According to this dynamical rule, the two first moments of the opinion distribution, i.e., the total mass and the mean opinion, are conserved. In particular, if we define the $k$th moment as $M_k(t) = \int dx \,\, x_k P(t,x)$, then it can be easily found that $\dot{M_0} = 0 = \dot{M_1}$. For $\epsilon \geq 0.5$ it is shown in \cite{BenNaim2003} that a stationary solution is found asymptotically for $t \to \infty$ as $P_{st} (x) = \delta (x - x_0)$, where $x_0$ is the mean opinion, that is, only one opinion cluster is formed. On the contrary, for lower values of the bound of confidence $\epsilon < 0.5$ it is found by simulation that a finite number of opinion groups are formed, leading to a stationary distribution as $P_{st} (x) = \sum_{i=1}^r m_i \delta (x - x_i)$, where $r$ is the number of clusters, and $x_i$, $m_i$ are respectively the position and the mass of cluster $i$.

We will follow here a slightly different approach, as developed in \cite{Lorenz2005Pro}, \cite{Lorenz2007PHD}, \cite{Lorenz2010}, involving a previous discretization of the opinion space into a finite number of opinion classes. The behavioral rules for the individuals are then translated to dynamical rules for the density of agents in these resulting opinion classes. Concerning the discretization, we divide the opinion space $[0, 1]$ into $n$ subintervals or \textit{opinion classes} as $\left[0, \frac{1}{n}\right)$, $\left[\frac{1}{n}, \frac{2}{n}\right)$, ... , $\left[\frac{n-1}{n}, 1\right]$, which are labeled as ${1,\dots,n}$. The bound of confidence $\epsilon$ is also naturally transformed into its opinion classes counterpart $[n \epsilon]$, where the square brackets $[]$ denote the integer part of a number. Therefore, the state of the system can be quantified by a vector $p(t)$, where each component $p_i (t)$ is the fraction of the total population which holds opinions in class $i$, and the time evolution of the system can be written as a discrete Markov chain. The density-based dynamics can then be defined as
\begin{equation}
 p(t+1) = p(t) T[p(t)],
\label{density-based}
\end{equation}
where $T(p(t))$ is a transition matrix depending only on the current state of the system $p(t)$, and which in the case of DW \textit{et al.} is
\begin{equation}
 T_{ij} [p] = \begin{cases}
               \frac{\pi_{2j-i-1}^i}{2} + \pi_{2j-i}^i + \frac{\pi_{2j-i+1}^i}{2}, & \text{if } i \neq j,\\
               q_i, & \text{if } i = j,
              \end{cases}
\label{transMatrix}
\end{equation}
with $\displaystyle q_i = \sum_{j \neq i, j=1}^n T_{ij} (p)$ and
\begin{equation}
 \pi_m^i = \begin{cases}
               p_m, & \text{if } |i - m| \leq [n \epsilon],\\
               q_i, & \text{otherwise}.
              \end{cases}
\label{pi}
\end{equation}
This transition matrix states that the probability of an agent to change from opinion $i$ to opinion $j$ depends only on the fractions of agents in the opinion classes $2j - i - 1$, $2j - i$ and $2j - i + 1$, and only if these classes are not farther than $[n\epsilon]$ from $i$. This is true because they are the only opinion classes whose average with $i$ results in $j$. In the case of $2j - i - 1$ and $2j - i + 1$ it is only half of the agents who jump to $j$, the other half jumping to $j - 1$ and $j + 1$ respectively. This discrete time and discrete opinion approach has been shown to lead to the same results as the previously presented continuous density-based model \cite{Lorenz2007PHD}.

Within this context, the knowledge of $P(0)$ implies the knowledge of the opinion distribution $P(t)$ at any point in time, avoiding the need of different realizations of the process. This is related to the fact that they correspond to the limit $N \to \infty$, very large number of agents, and thus to a case without finite-size fluctuations. However, we must stress that the differences between the agent and the density-based approaches will therefore be greater the smaller the number of agents \cite{Raul2009}. Some examples of disagreement will be commented below. Another interesting advantage of the density-based approaches is that the conservation of the symmetry can be directly observed, in the continuous case, in the master equation \eqref{originalME} and, in the discrete case, in the definition of the transition matrix. This latter point implies that, if the initial distribution of opinions is symmetric around the central point $x = \nicefrac{1}{2}$, then we will have $\forall t$ that $P(t,x) = P(t, 1 - x)$.

In order to determine the attractive cluster patterns for each bound of confidence and observe transitions of attractive patterns at critical values of $\epsilon$, bifurcation diagrams are drawn. The asymptotic results of the model are shown in these graphs, where the location of clusters is plotted versus the continuum of values of the bound of confidence $\epsilon$. \textit{Bifurcation} refers here to the appearance, dominance or splitting of a given cluster. Sometimes it is useful, when studying cluster patterns, to establish a difference between ``major'' opinion clusters, containing a high fraction of the population, and ``minor'' opinion clusters, containing a much smaller fraction. The bifurcation diagram for the DW \textit{et al.} model with uniform initial density in the opinion space is shown in the plot \textbf{b} of figure \ref{bifurQuad}. A detailed analysis \cite{BenNaim2003} shows that there are in these results four basic modes of bifurcation, which are repeated in descendant order of the bound of confidence $\epsilon$ and in shorter and shorter $\epsilon\textrm{-intervals}$. Let us describe, as an example, the first sequence. For $\epsilon \geq 0.5$ only one big central cluster evolves, gathering the vast majority of the population. As the bound of confidence decreases from $\epsilon = 0.5$, we can notice the nucleation of two minor clusters from the boundaries of the opinion interval. Around $\epsilon \approx 0.266$ there is a bifurcation of the central cluster into two major clusters. Further decreasing $\epsilon$, the central cluster has a rebirth as a minor cluster from $\epsilon \approx 0.222$, before suddenly increasing its mass around $\epsilon \approx 0.182$, pushing the two major clusters outwards.

A topic that has received much less attention in the relevant literature is the dependence of the model on the initial conditions, that is, on the initial distribution of opinions among the agents. As a first and naive argument, we could notice that the model, as has been presented, conserves the mass and the mean opinion of the population, so the position of the final clusters will depend on the mean of the initial distribution. However, also different initial conditions with the same mean opinion could give rise to different final configurations. In fact, it can be shown theoretically and confirmed by simulation that any combination of delta-functions is a steady state solution of the master equation describing the model, provided these delta-peaks are separated by a distance greater than $\epsilon$ and they conform to the mean opinion conservation \cite{Lorenz2007PHD}, \cite{Raul2009}. This will be the object of study in section \ref{originalRes}, where we will show that it is perfectly possible to force or prevent a consensus by varying the initial distribution of opinions.


\subsection{The model with noise}
\label{noisyMod}

A final configuration consisting of one or more delta functions means that all the agents within one of these opinion groups would share exactly the same opinion, which is clearly not very realistic. In order to avoid this perfect consensus within each cluster, Pineda, Toral and Hern\'andez-Garc\'ia presented an extension of the original DW \textit{et al.} model taking into account some additional randomness \cite{Raul2009}. Thus, a noise is introduced into the dynamics as a \textit{free will}, which allows for the agents to change their opinion from time to time to a randomly chosen value. This is equivalent to allowing each agent to go back to a certain opinion preferred by this agent, that we will call \textit{basal distribution}. Thus, we implement a quenched noise, different from some other adaptive noise approaches \cite{Helbing2010}.

The dynamics is therefore modified by allowing, at each time step, a randomly chosen agent to follow the original DW \textit{et al.} interaction rule with probability $(1 - m)$ or to return to its preferred, basal, opinion with probability $m$. The probability $m$ is therefore a measure of the noise intensity $m_c$. As a result of this extension, there exists a transition between an organized and a disorganized phase at a critical value of the noise intensity. In the disordered state, corresponding to quite high noise rates, there is no cluster formation, as noise is stronger than nucleation processes; on the other hand, in the ordered state, corresponding to lower noise rates, we can still clearly observe the formation of clusters, even if they broaden with respect to the noiseless case. Another difference between the original and the noisy models is that the position of the clusters in the latter case only vary at the bifurcation points, remaining constant between them. These features are shown in the plot \textbf{b} of figure \ref{NoiBifurQuad}, where the asymptotic probability distribution $P_{st}(x)$ is plotted as a function of the bound of confidence $\epsilon$ for a small noise intensity $m=0.01$.

Taking as a starting point the master equation of the original DW \textit{et al.} model, \eqref{originalME}, the dynamics of the noisy case can be derived as
\begin{equation}
 \begin{split}
 \frac{\partial}{\partial t} P(t,x) = (1 - m) \int_0^1 dx_1 \int_{|x_1-x_2| \leq \epsilon} dx_2 \Bigg[ P(t,x_1) P(t,x_2) \bigg( 2 \delta \left( x - \frac{x_1 + x_2}{2} \right)\\
 - (\delta (x - x_1) + \delta(x - x_2)) \bigg) \Bigg] \; + \; m (\underbrace{P_a (x) - P(t,x)}_\text{Noise term}),
 \end{split}
\label{noisyME}
\end{equation}
where the $P_a (x)$ is the distribution of preferred opinions or the probability distribution of the noise. It is clear from equation \eqref{noisyME} that the net effect of the noise is to move the current opinion distribution towards the preferred opinions distribution with a velocity or intensity per time step $m$.

An interesting feature of this extension is that the average opinion of the system is not anymore a constant and it tends to the average of the noise distribution, regardless of that of the initial condition. The time evolution of the first moment will thus be
\begin{displaymath}
 \frac{dM_1}{dt} = m [ M_1^a - M_1 ],
\end{displaymath}
where $M_1^a$ is the first moment of the distribution $P_a (x)$.

We devote section \ref{noisyRes} to the study of the influence of the initial distribution of opinions in this noisy case. There, we will show that, even if the importance of the initial condition regarding the average opinion is replaced by that of the noise distribution, the former has still some influence in determining the bifurcation patterns. 



\section{Initial conditions, measures and simulations}
\label{measures}

We present in this section the two sets of different initial conditions that we have employed in our simulations, as well as the measure used to characterize and differentiate between these distributions. Furthermore, we also present in this section some measures developed in the literature for the analysis of the resulting final configurations. Finally, we also give some details about the simulations performed.

\begin{figure}[ht!]
\centering
\includegraphics[width=!, height=7cm]{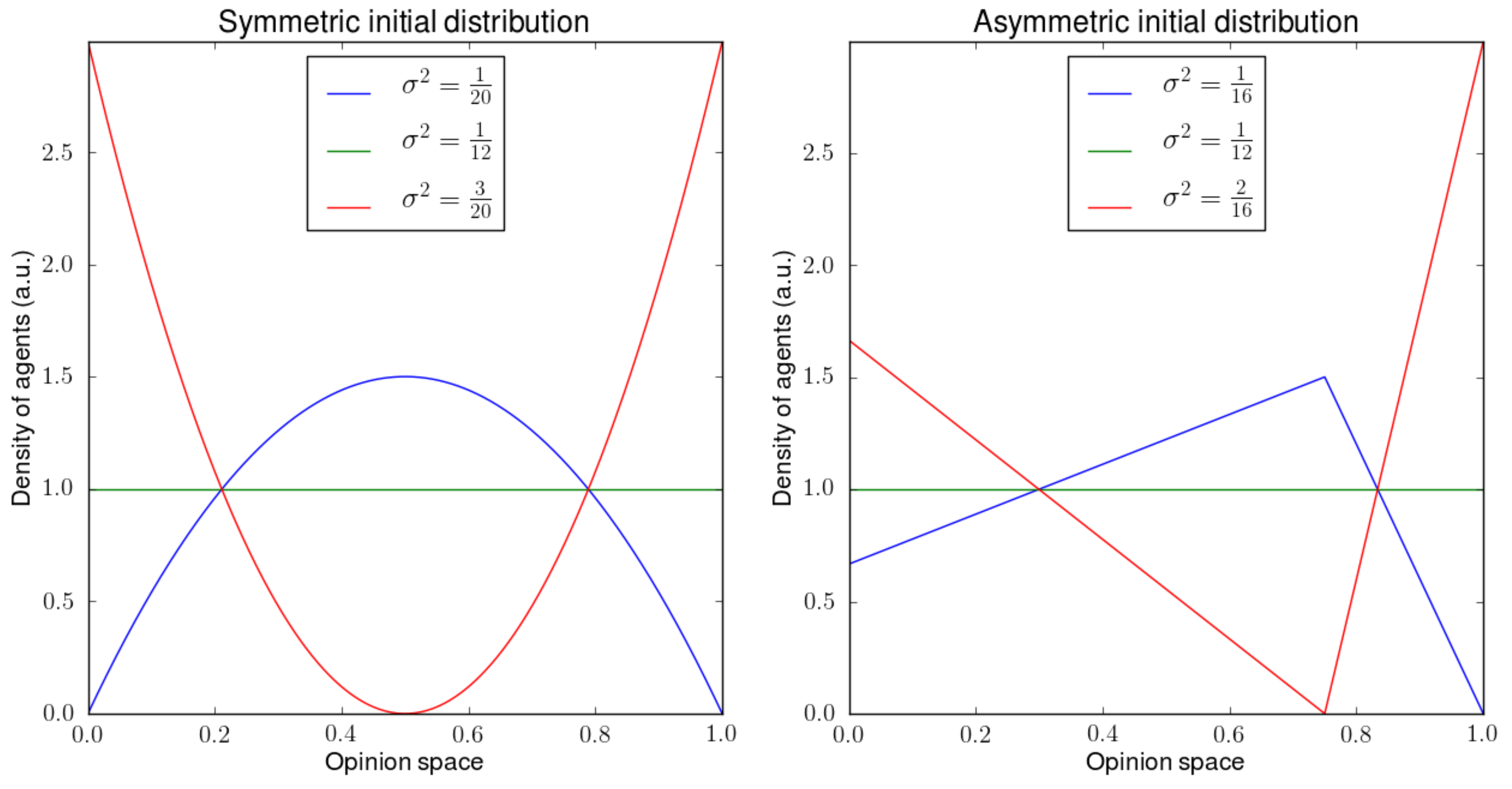}
\caption{\label{initialDist} Examples of the symmetric and asymmetric sets of initial distributions of opinions.}
\end{figure}

Regarding the initial conditions, we used basically two different sets of distributions. First, we defined a set of symmetric initial distributions, quadratic functions in particular, as
\begin{equation}
 P(0,x) = 1 + b \left[ \left( x - \frac{1}{2} \right)^2 -\frac{1}{12} \right],
\label{quadratic}
\end{equation}
where the different coefficients have been tuned so that, by varying the parameter $b$ in the range $b \in [-6, 12]$, the quadratic function smoothly changes from a concave to a convex shape at $b = 0$. The reason to set this particular interval for $b$ is that the density of agents in the different opinion classes remains non-negative. All of these initial conditions are symmetric and correspond to an average opinion of $\nicefrac{1}{2}$. Afterwards, we also tested the effect of avoiding this symmetry but keeping the average opinion of $\nicefrac{1}{2}$. As it is not possible to do so with a quadratic function, we defined a set of asymmetric ``triangular'' distributions as

\begin{equation}
P(0,x) = 
\begin{cases}
\displaystyle
\vspace{5pt}
1 + a\left(\frac{3}{10}-x\right)  & \text{if } x \leq \frac{3}{4},\\
\displaystyle
1 + \frac{27}{5} a\left(x- \frac{5}{6}  \right) & \text{if } x > \frac{3}{4},
\end{cases}
\label{triangular}
\end{equation}
where the parameter $a$, restricted to $a \in [\nicefrac{-10}{9}, \, \nicefrac{20}{9}]$, sets the concavity or convexity of the complete function. Both sets of distributions are exemplified in figure \ref{initialDist} for the maximum and minimum values of the respective $a$ and $b$ intervals.

Although this is not the only possibility, and in order to characterize and compare the different initial conditions, we decided to use the ``shifted variance" defined as  $S_2=\sigma^2-\nicefrac{1}{12}$, being $\sigma^2$ the variance, or second central moment of the distribution around the mean. As $S_2=\nicefrac{b}{180}$ for the quadratic distribution and $S_2=\nicefrac{3a}{160}$ for the triangular condition, it correctly distinguishes situations where a majority of agents have initially opinions around the center of the opinion interval ($a,b>0$) from other cases where a majority of agents have opinions towards the extremes ($a,b<0$). The former corresponds to an initial condition which would favor consensus while the latter corresponds to facilitating the splitting of the population into, at least, two opinion groups.

Since we are interested on assessing the effect of the initial condition on the final state of the system, we use some measures of consensus already introduced in the literature to quantify this final distribution of opinions. In particular, we will use bifurcation diagrams as already described in section \ref{originalMod}. For the construction of these diagrams, we need a precise definition of what a cluster is and what its position in the opinion space is \cite{Lorenz2007PHD}. We define a \textit{cluster} with a given precision, in our case $10^{-3}$, as a set of adjacent opinion classes each of them with a fraction of agents or mass greater than the precision threshold and surrounded by neighboring classes with masses less than this precision. Regarding the \textit{position} in the opinion space or the opinion corresponding to a given cluster, we approximate it by the average opinion of the agents within this group. In order to quantify the degree of consensus within the population, the \textit{mass of the biggest cluster} has been pointed out as an appropriate measure. We define the mass of the biggest cluster as the fraction of agents within the largest of the clusters \cite{Lorenz2010}.

As it will be explained below, the introduction of a noise in the original DW \textit{et al.} model allows the system to undergo certain transitions between states which were forbidden in the original model. We could argue, in a very naive manner, that this effect is related to the fact that these states are stable or metastable, i.e., to their relative stability. Thus, only in the presence of a noise would a transition from a metastable to a globally stable state be allowed. In order to characterize the relative stability of those states we defined a Lyapunov function for the DW \textit{et al.} model, that is, a positive-definite function which always decreases in time, given the interaction rules of the model, and whose minimum is zero. Let us write here only the definition of the Lyapunov function, while we leave its proof for the appendix \ref{appendix},
\begin{equation}
 \mathcal{L}[\vec{x}] = \sum_{i > j} (x_i - x_j)^2,
\label{Lyapunov}
\end{equation}
where $\vec{x}(t)$ is a vector whose components are the opinions of the agents at time $t$. In the absence of noise, only transitions from states with a given value of $\mathcal{L}$ to those with a smaller (or equal) value would be permitted:
\begin{displaymath}
 \mathcal{L}[\vec{x}(t+1)] \leq \mathcal{L}[\vec{x}(t)].
\end{displaymath}

Concerning the characteristics of the numerical simulations, and as we have already pointed out in section \ref{introduction}, we employed an algorithmic approach based on discrete density-based reformulation of the DW \textit{et al.} model developed by J. Lorenz \cite{Lorenz2005Pro}, \cite{Lorenz2007PHD}. In particular, we discretized the opinion space $[0, 1]$ into $1001$ opinion classes, an odd number being the only option allowing for one-class central opinion clusters. This one-class central opinion group in class $501$ is quite important since, in the symmetric case, it is the only class which can have a density greater than $0.5$. In order to study the bifurcation patterns, we run the algorithm for $100$ evenly distributed values of the bound of confidence parameter, from $\epsilon = 0.1$ to $\epsilon = 0.6$. Furthermore, and with the purpose of analyzing the influence of the initial condition, we performed the simulations for $100$ different values of the initial distribution parameters $a$ and $b$, for each value of the bound of confidence.

In the noisy extension of the DW \textit{et al.} model, we set the intensity of the noise at $m = 0.01$ in our simulations, in order for the system to be in the ``ordered'' state, as explained in subsection \ref{noisyMod} and in \cite{Raul2009}. The convergence time, that is, the number of time steps needed for the system to arrive at a converged state with opinion clusters separated by more than a bound of confidence, is much longer in the noisy than in the noiseless case. Therefore, the $1000$ time steps we used for the original model were more than enough in that case, while we decided to use $50000$ for the noisy model in order to ensure a good level of convergence.

\section{The importance of the initial conditions in the noiseless model}
\label{originalRes}

In the noiseless case, when $m = 0$, the asymptotic steady state distribution $P_{st}(x) = \lim_{t \to \infty} P(x,t)$ is basically a sum of delta-functions located at certain points. As we are only interested in the fundamental changes caused by the variation of the initial conditions, we have chosen a relatively high threshold level for the detection of opinion groups, $10^{-3}$, in order not to detect the minority clusters but only the major ones.

\begin{figure}[ht!]
\centering
\includegraphics[width=!, height=9cm]{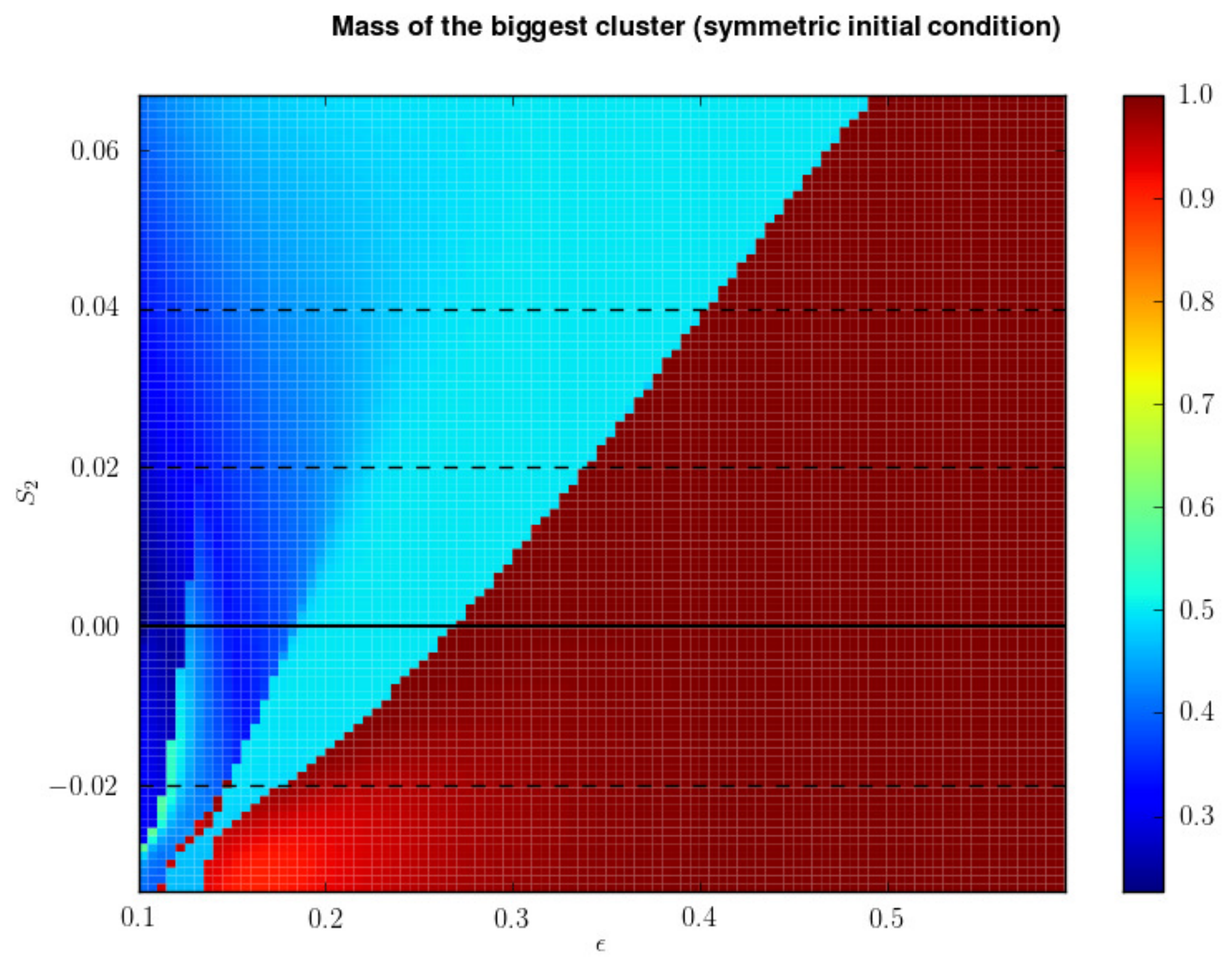}
\caption{\label{MassQuad} Mass of the biggest cluster for the set of symmetric initial distributions of opinions (quadratic functions) obtained with the noiseless DW \textit{et al.} model. The black horizontal lines mark the values of the shifted variance for which cuts of the opinion distribution are shown in figure \ref{bifurQuad}.}
\end{figure}

\begin{figure}[ht!]
\centering
\includegraphics[width=!, height=9cm]{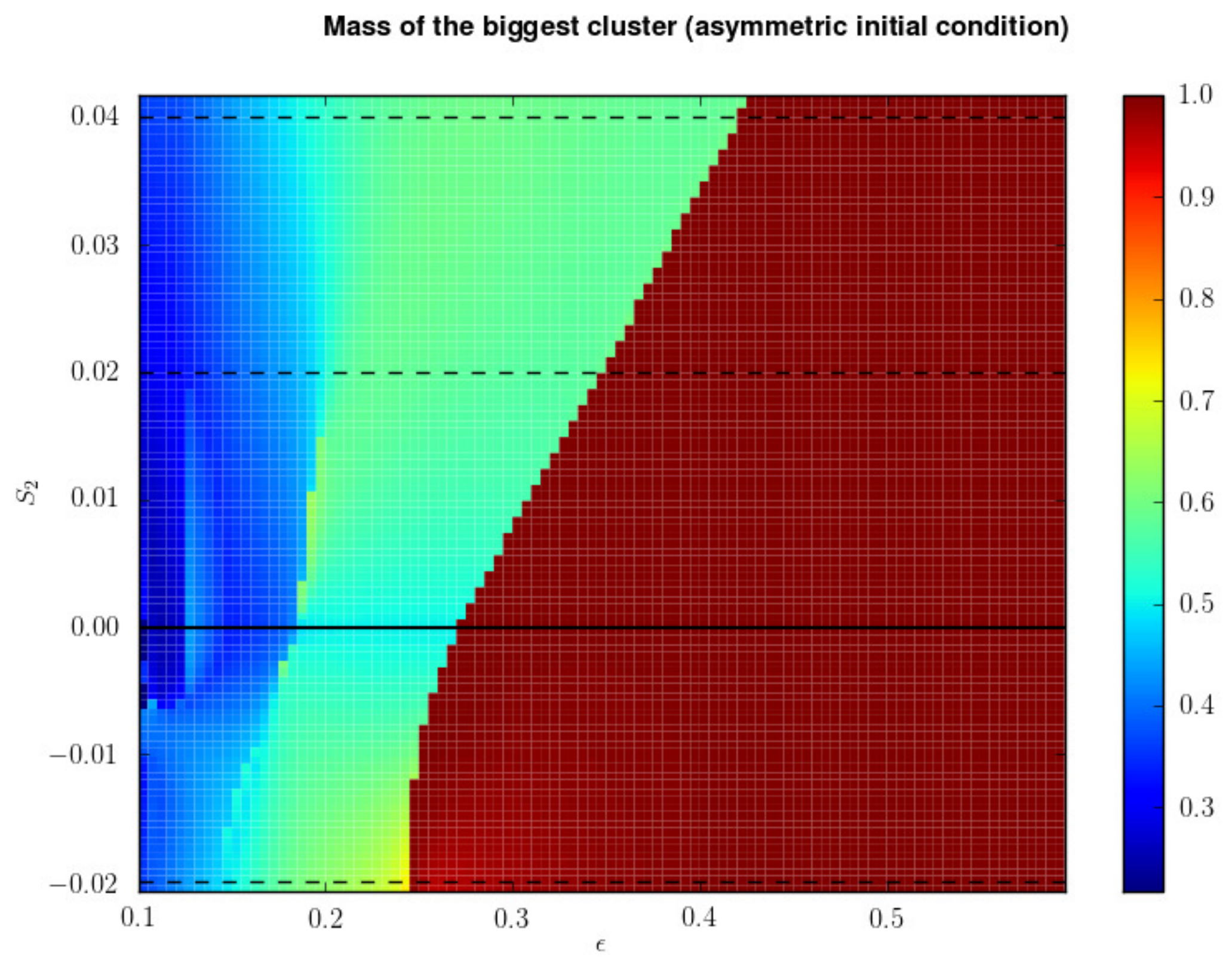}
\caption{\label{MassTri} Mass of the biggest cluster for the set of asymmetric initial distributions of opinions (triangular functions) obtained with the noiseless DW \textit{et al.} model. The black horizontal lines mark the values of the shifted variance for which cuts of the opinion distribution are shown in figure \ref{bifurTri}.}
\end{figure}

\begin{figure}[ht!]
\centering
\includegraphics[width=14cm, height=!]{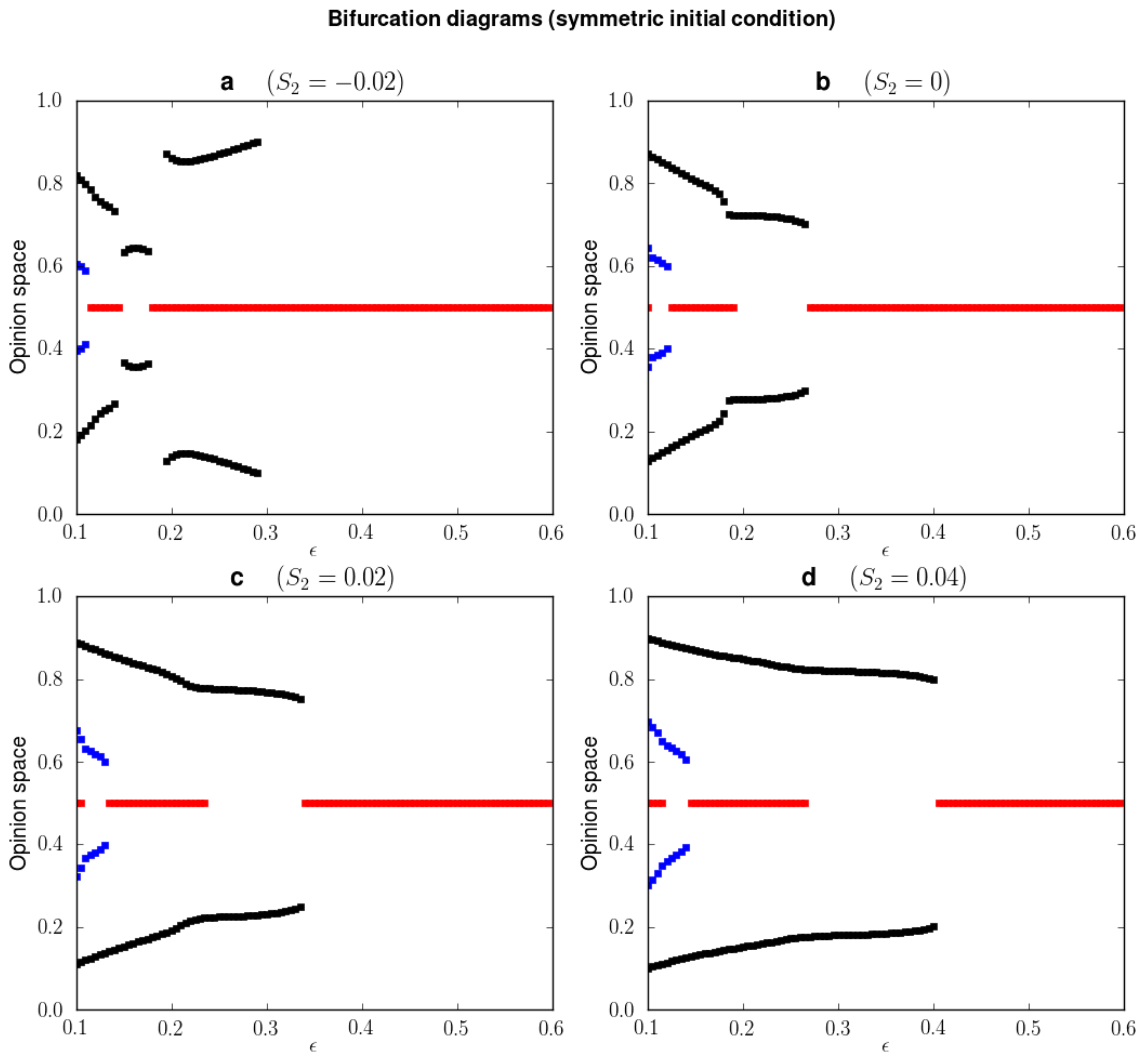}
\caption{\label{bifurQuad} Bifurcation diagrams for four different symmetric initial distributions of opinions (quadratic functions) obtained with the noiseless DW \textit{et al.} model.}
\end{figure}

\begin{figure}[ht!]
\centering
\includegraphics[width=14cm, height=!]{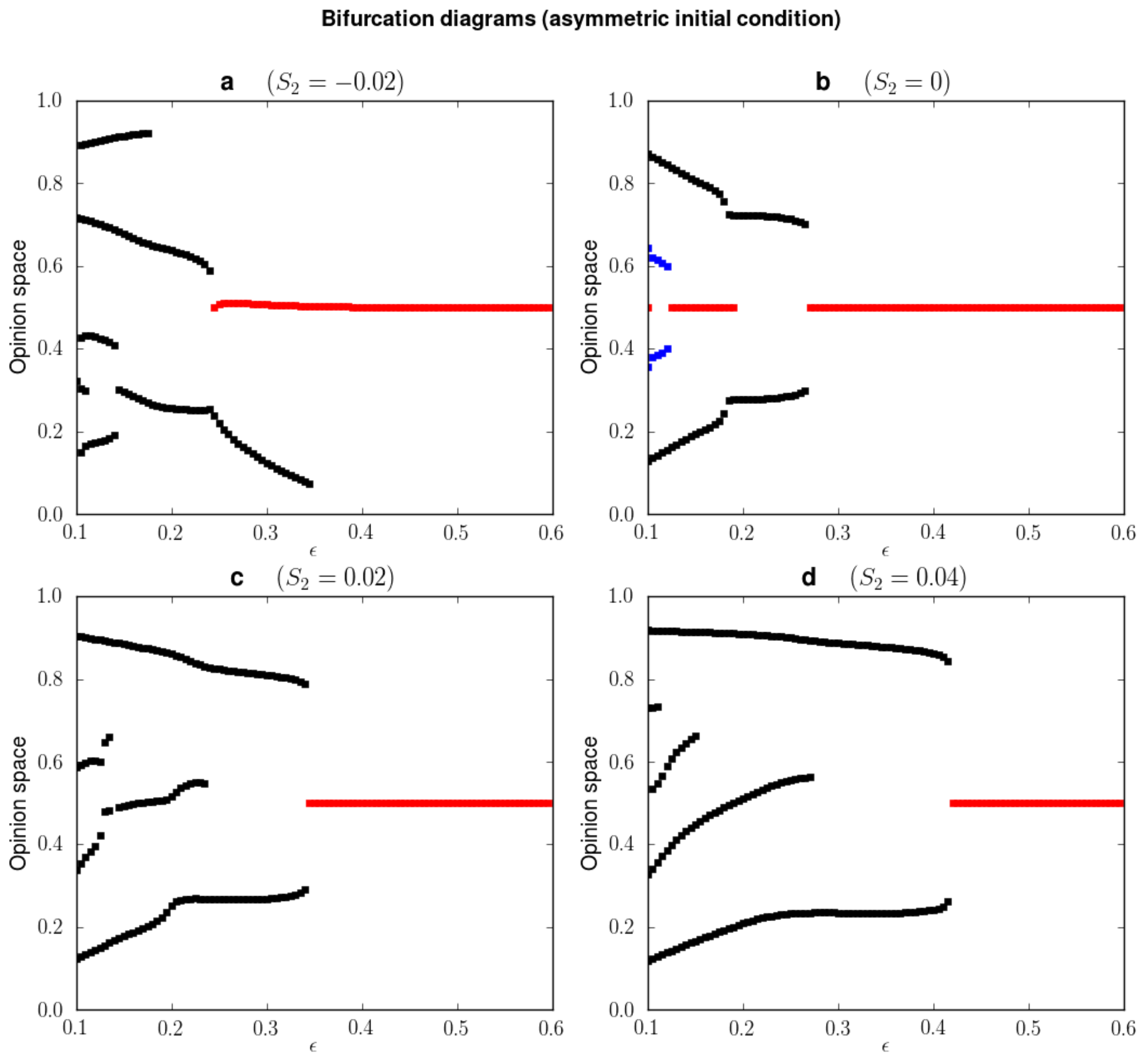}
\caption{\label{bifurTri} Bifurcation diagrams for four different asymmetric initial distributions of opinions (triangular functions) obtained with the noiseless DW \textit{et al.} model.}
\end{figure}

The masses of the biggest cluster can be observed in figures \ref{MassQuad} and \ref{MassTri} for the symmetric and the asymmetric initial condition cases respectively. In these plots, the x-axis represents the values of the bound of confidence $\epsilon$, while the y-axis corresponds to the shifted variance $S_2$. Each point of this plane $(\epsilon, \, S_2)$ is colored according to the values of the mass of the biggest cluster after stabilization. Parameter regions with major consensus are therefore colored dark red, while light blue means two equally distributed clusters. In order to give a more exhaustive idea about the bifurcation patterns which give rise to the masses of the biggest cluster shown above, we present in figure \ref{bifurQuad} the bifurcation diagrams for the four cuts or horizontal lines marked in figure \ref{MassQuad}, and in figure \ref{bifurTri} for those marked in figure \ref{MassTri}.

In figures \ref{MassQuad} and \ref{MassTri} we see some clear and sharp lines of transition, while some others look like more smooth and gradual changes. Let us first focus on the symmetric case, figure \ref{MassQuad}, where we observe an oblique line of sudden transition between dark red and light blue, which can be shown to mean (see figure \ref{bifurQuad} below) a bifurcation from one central major opinion cluster to two smaller and symmetrical clusters. The smooth transition between light and dark blue represents a gradual decrease of the mass of these symmetrical clusters and the rebirth of the central one. In any case, the point we would like to stress here is the fact that these transition lines or regions, their position and smoothness, are highly dependent on the variance of the initial distribution of opinions, something that is particularly evident for the sharp transition from consensus to more than one opinion cluster.

Focusing now on the asymmetric initial condition case, we notice in figure \ref{MassTri} a rather different transition pattern. The main difference is the appearance of a new color or state, turquoise blue (between green and blue), meaning a mass of around $0.6$ but corresponding to a state with two major opinion clusters, which would be impossible in the previous and symmetrical case. This is due to the fact that the initial distribution is not anymore symmetric around the central opinion $0.5$, so a cluster can exist with a mass larger than $0.5$ and not being located at the center of the opinion space. Since, when the shifted variance equals zero, both initial conditions are equal and uniform, the black solid horizontal line is identical in both figures.

Four examples of bifurcation diagrams are presented in figures \ref{bifurQuad} and \ref{bifurTri} for each set of initial conditions, corresponding to the cuts marked in figures \ref{MassQuad} and \ref{MassTri} respectively. By comparison between these bifurcation diagrams and the previous plots of the mass of the biggest cluster, we can notice that some of the opinion groups found are actually minority clusters. They are detected in the bifurcation diagram but, since they have a mass just slightly larger than the precision threshold $10^{-3}$, they have no influence in the mass of the biggest cluster. The most clear examples of these minority groups are, on the one hand, the two symmetric clusters appearing on the bifurcation diagram for $S_2 = -0.02$ in figure \ref{bifurQuad} from $\epsilon \approx 0.3$ to $\epsilon \approx 0.2$ and, on the other hand, the cluster appearing just from the lower end of the opinion interval from $\epsilon \approx 0.35$ to $\epsilon \approx 0.25$ again on the bifurcation diagram for $S_2 = -0.02$ in figure \ref{bifurTri}.

Both figures \ref{bifurQuad} and \ref{bifurTri} serve as a confirmation of the strong importance of the initial condition in determining the bifurcation patterns of the steady state solution in the noiseless DW \textit{et al.} model. In the symmetric case, even though the pattern of bifurcations is rather similar in each and every example, the points of the bound of confidence axis where the bifurcations do take place smoothly increase with the variance of the initial distribution of opinions. This change is particularly relevant for the first bifurcation, which takes place at $\epsilon \approx 0.270$ in the uniform case, shown in plot \textbf{b}, but moves from $\epsilon \approx 0.180$ in plot \textbf{a} to $\epsilon \approx 0.405$ in plot \textbf{d}. Regarding the case of asymmetric initial conditions shown in figure \ref{bifurTri}, it is not only the bifurcation points, but also the general bifurcation structure which is perturbed by a variation of the initial distribution of opinions. In particular, we observe that the bifurcation patterns shown in plots \textbf{a}, \textbf{c} and \textbf{d} are clearly asymmetric and the first bifurcation point moves from $\epsilon \approx 0.245$ to $\epsilon \approx 0.420$.

In order to check whether the particular shape of the initial condition function has any important influence upon the bifurcation patterns of the model, we also tested a symmetric initial distribution with the shape of a centered triangle, thus different from the quadratic one. The results (not shown) obtained being remarkably similar to the case of the symmetric quadratic condition, we deduce that the particular shape of the distribution does not play a major role regarding the final asymptotic solution of the model. Therefore, we notice that the two most important variables for the characterization of the initial condition and its effects upon the DW \textit{et al.} model are the symmetry and the variance of the distribution.

\section{The importance of the initial conditions in the noisy model}
\label{noisyRes}

If a noise of the type described in subsection \ref{noisyMod} is added to the original system, that is, if $m > 0$, then the asymptotic solution of the model is not anymore a collection of delta-functions, but a smooth distribution of opinions. Furthermore, if this noise $m$ is small enough for the system to be in the ordered state, the smooth steady state solutions will be peaked around some particular values. In particular, we set the noise intensity at $m = 0.01$. Although, as a consequence, the definition of a cluster is not so obvious in this case as in the original noiseless model, the cluster detection mechanism presented in section \ref{measures} aims to give correct results also in this noisy situation, and so we will use it for the mass of the biggest cluster plots. However, with regard to the bifurcation diagrams, and following \cite{Raul2009}, we decided to show here the whole probability density distribution and not just the position of the major clusters.

\begin{figure}[ht!]
\centering
\includegraphics[width=!, height=9cm]{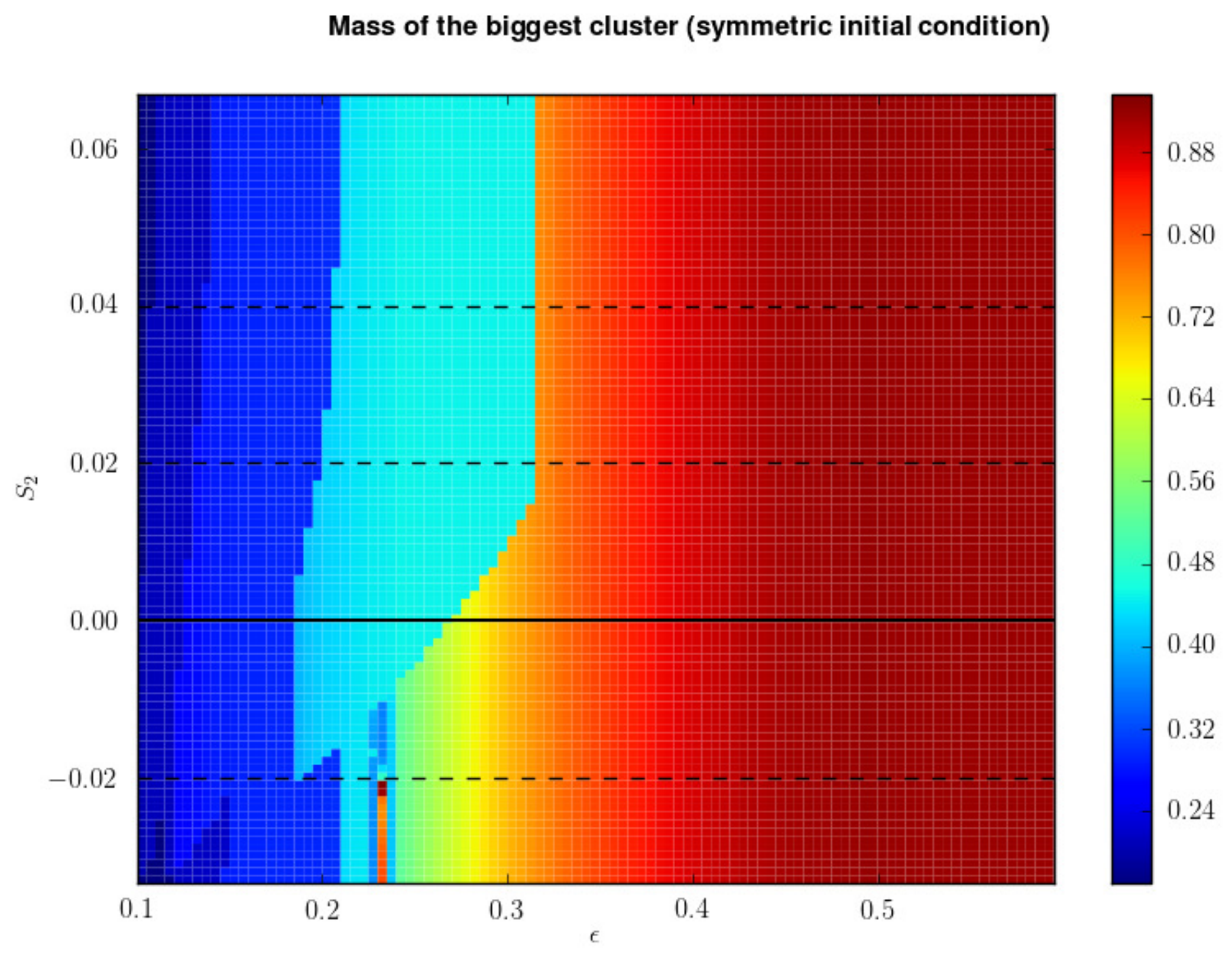}
\caption{\label{NoiMassQuad} Mass of the biggest cluster for the set of symmetric initial distributions of opinions (quadratic functions) obtained with the noisy DW \textit{et al.} model with a noise intensity $m = 0.01$. The black horizontal lines mark the values of the shifted variance for which cuts of the opinion distribution are shown in figure \ref{NoiBifurQuad}.}
\end{figure}

\begin{figure}[ht!]
\centering
\includegraphics[width=!, height=9cm]{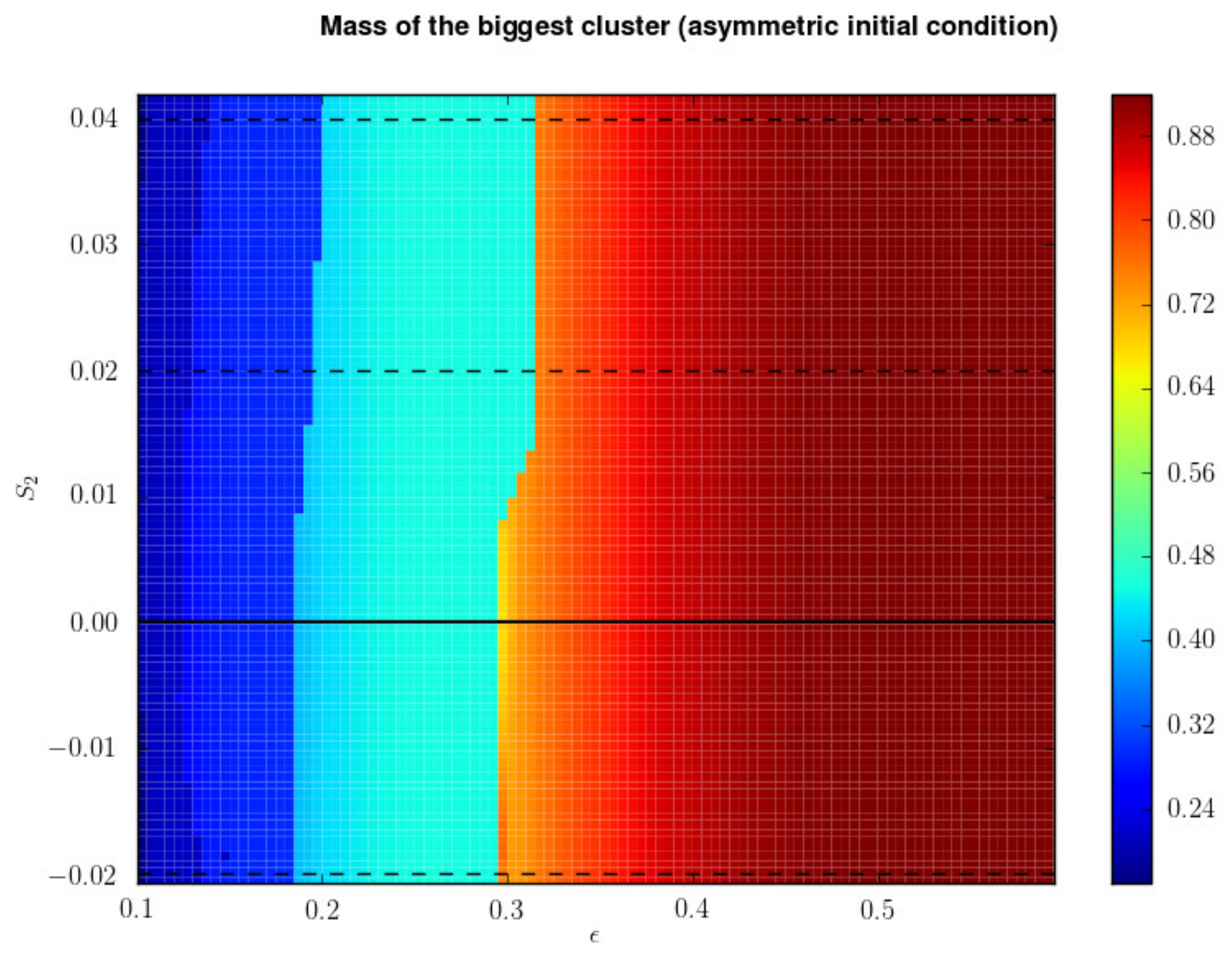}
\caption{\label{NoiMassTri} Mass of the biggest cluster for the set of asymmetric initial distributions of opinions (triangular functions) obtained with the noisy DW \textit{et al.} model with a noise intensity $m = 0.01$. The black horizontal lines mark the values of the shifted variance for which cuts of the opinion distribution are shown in figure \ref{NoiBifurTri}}
\end{figure}

\begin{figure}[ht!]
\centering
\includegraphics[width=14cm, height=!]{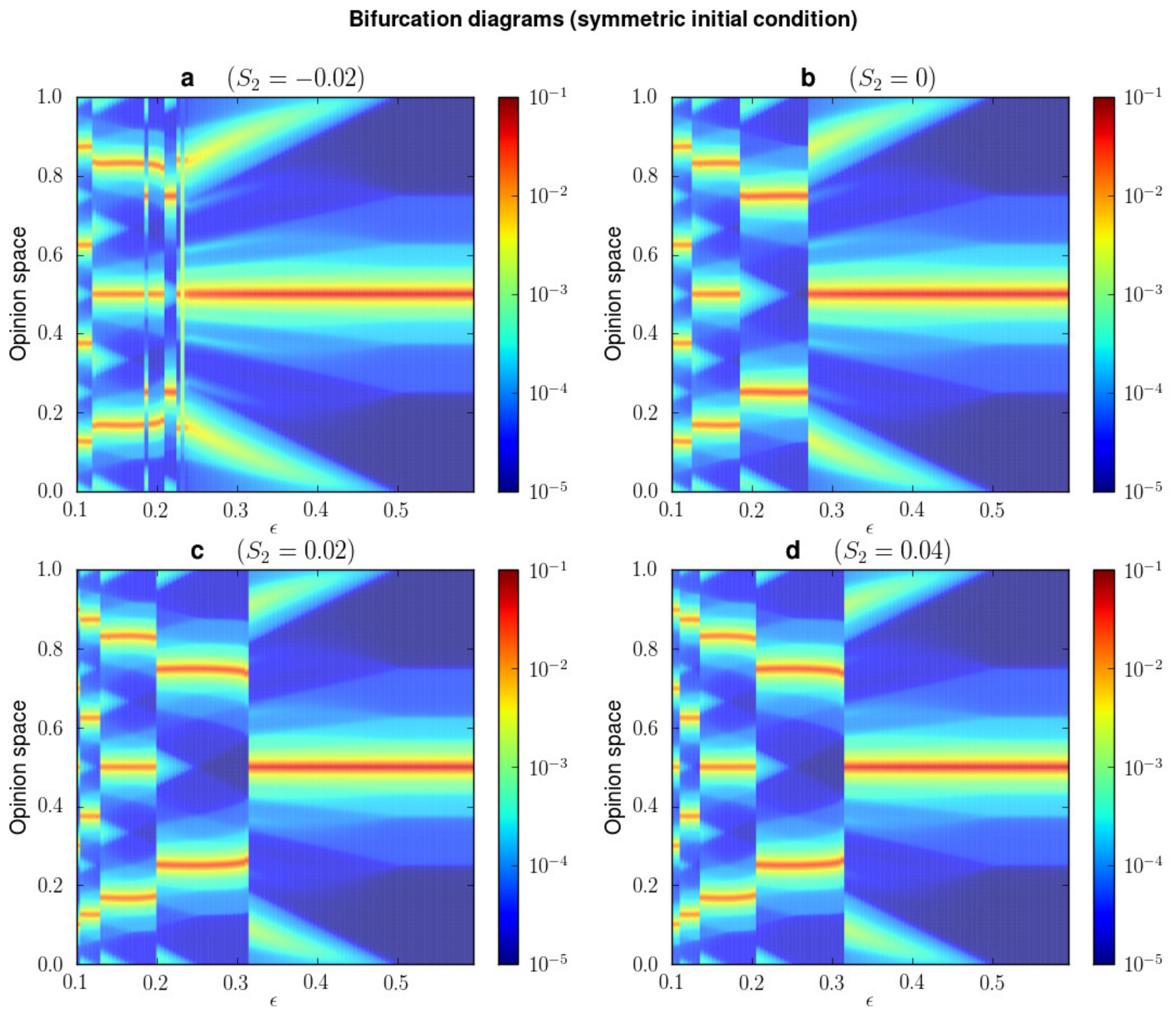}
\caption{\label{NoiBifurQuad} Bifurcation diagrams for four different symmetric initial distributions of opinions (quadratic functions) obtained with the noisy DW \textit{et al.} model with a noise intensity $m = 0.01$.}
\end{figure}

\begin{figure}[ht!]
\centering
\includegraphics[width=14cm, height=!]{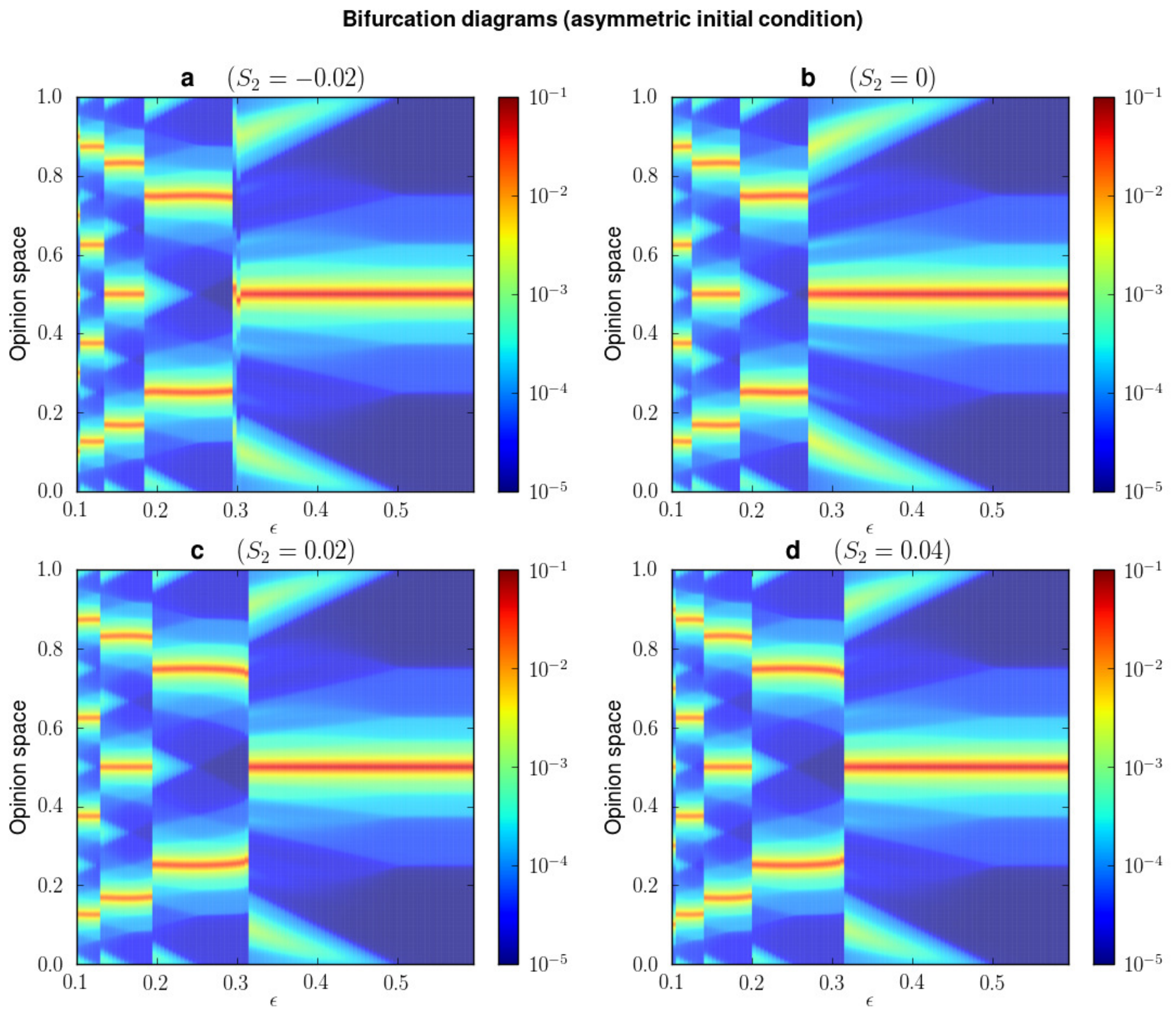}
\caption{\label{NoiBifurTri} Bifurcation diagrams for four different asymmetric initial distributions of opinions (triangular functions) obtained with the noisy DW \textit{et al.} model with a noise intensity $m = 0.01$.}
\end{figure}

As in the previous subsection, the mass of the biggest cluster is shown, for the noisy model, in figures \ref{NoiMassQuad} and \ref{NoiMassTri} for the symmetric and asymmetric initial condition cases respectively. It is interesting to notice, in a first and general view, that the range of values is slightly shorter in the noisy case than in the noiseless results. This is mainly due to the fact that the agents no longer gather in just a small number of opinion classes, but they are distributed all along the opinion interval. Therefore, even if the distribution is peaked around some popular opinion classes, a non negligible fraction of the population is dispersed among the rest of the opinion classes, not taking part in any major opinion cluster. This effect is also clearly observed in figures \ref{NoiBifurQuad} and \ref{NoiBifurTri}, which show a logarithmic colormap of the density of agents in each opinion class for the four cuts marked, respectively, in figures \ref{NoiMassQuad} and \ref{NoiMassTri}. These latter plots are, as already pointed out, the equivalent of the bifurcation diagrams of section \ref{originalRes}.

As before, we can observe in the mass of the biggest cluster plots of figures \ref{NoiMassQuad} and \ref{NoiMassTri} some sharp as well as some smooth lines of transition. Comparing with the noiseless case results in figures \ref{MassQuad} and \ref{MassTri}, we may notice that the transition lines are now mostly straight and vertical unlike the oblique lines observed before. This feature basically means that the initial conditions are, in this case, less important in determining the final configuration of the system. However, we can observe a smaller but still noticeable influence in the oblique transition line from $S_2 = -0.02$ to $S_2 = 0.02$ in figure \ref{NoiMassQuad}, and in the small change of the transition line around $S_2 = 0.01$ in figure \ref{NoiMassTri}.

We know, by computing the value of the Lyapunov function \eqref{Lyapunov}, that the most stable situation ---with $\mathcal{L} = 0$, in particular--- is that where all the agents share the same opinion, and thus form only one cluster. This is the most stable configuration, in the sense of the Lyapunov function defined before, even if there is a certain small spread in the opinion of the group participants. A configuration with two clusters corresponds to much higher values of the Lyapunov function, but still it may constitute a metastable state when the bound of confidence does not allow for interactions between the agents of both groups. Therefore, when the system starts from an initial condition very near to the two clusters final state, then it quickly evolves towards that final configuration. However, when introducing a small noise in this last situation, then the system is able to overcome the barrier introduced by the bound of confidence and undergo a transition to the globally stable state of one cluster. This is so for large and intermediate values of the bound of confidence, for $\epsilon > 0.310$. On the contrary, for small values of this threshold, the time needed to arrive at a consensus becomes so large relative to the noise rate that the system is not able to leave the two clusters state. Regarding those initial conditions imposing a certain consensus from the beginning, even if their initial state is already more stable than a two cluster configuration, the presence of noise is able, for small values of the bound of confidence, to take the system to the two opinion groups state, which is the metastable state most easily reached from the noise opinion distribution, uniform in our case. Jumps between the stable and the metastable configurations have indeed been reported in the case of a uniform initial condition \cite{Raul2009,Raul2011}.

We show in figures \ref{NoiBifurQuad} and \ref{NoiBifurTri} four examples of bifurcation diagrams, in the form of probability density distributions, for each set of initial conditions. In these plots we can clearly observe some clusters and bifurcation patterns similar to the those in the bifurcation diagrams of the previous noiseless case. However, we observe two significant differences, apart from the fact that now we have a smooth distribution of opinions in the final state and not anymore a collection of delta-functions. On the one hand, the bifurcation points are now very sharply defined, with no region of progressive transition from one state to the next. On the other hand, the location of the maximal values of the density does not significantly depend on the bound of confidence between bifurcation points, unlike the position of clusters in the noiseless model. This may be due to the fact that the noisy perturbation allows the system to always arrive at the most stable configuration permitted by the bound of confidence, which only changes at the bifurcation points, when this confidence distance allows for more clusters to appear. On the contrary, the noiseless model freezes on stable configurations even if they are not the most stable.

It is important to remark, in figures \ref{NoiBifurQuad} and \ref{NoiBifurTri}, the considerably long times needed for convergence when the initial conditions strongly force a consensus, as it is the case of the plot \textbf{a} in both figures, where $S_2 = -0.02$. There, we can see some blurs on the colormap showing that the system has not perfectly converged even after $50000$ time steps. Contrary to the results of the noiseless case, now the bifurcation patterns shown have exactly the same structure for all the initial distributions tested. Nevertheless, there is still a perceivable influence of the initial conditions, as the location of the bifurcation points clearly changes among the different plots of the bifurcation diagrams, confirming what was noted before about the mass of the biggest cluster plots. Excluding the first and unconverged plot \textbf{a}, the first bifurcation point, from consensus to two opinion clusters, moves from a minimum value of $\epsilon \approx 0.270$ in the uniform initial condition plot \textbf{b} to a maximum of $\epsilon \approx 0.315$ in plot \textbf{d} for both the symmetric and the asymmetric cases.

Another interesting feature to notice in the bifurcation diagrams is the strong difference between figure \ref{NoiBifurTri}, in this subsection, and \ref{bifurTri}, in the previous one. They both show the results of the model starting from an asymmetric initial condition. However, only in the noiseless case the resulting bifurcation diagram is asymmetric. The graph obtained with the noisy model for an asymmetric initial distribution is, in fact, equal to the one obtained with a symmetric initial condition. This is a case of symmetry restored by noise, and it is due to the fact that the distribution of the noise or preferred opinions we use is always symmetric and uniform. As it was proved in \cite{Raul2009} and shown in subsection \ref{noisyMod}, the prevalent condition, regarding the moments of the final distribution, is the basal condition of the noise or preferred opinions distribution of the agents. Thus, the mean opinion value to be conserved and the general bifurcation structure to be found is that corresponding to a uniform distribution.

Again, we checked that the particular shape of the initial distribution has no major effect concerning the asymptotic solution of the model by simulating also a symmetric initial distribution with the shape of a centered triangle, thus different from the quadratic one. As in the noiseless case, we found that the results are remarkably similar to the case of the symmetric quadratic condition. So we confirm that the symmetry and the variance of the distribution are still quite good parameter for the characterization of the initial condition and its effects upon the DW \textit{et al.} model.

%

\section{Conclusions}
\label{conclusions}

We have shown, by numerical simulation, that the asymptotic solution of the original DW \textit{et al.} model is highly dependent on the initial condition. As a consequence, we have shown that it is indeed possible to force or prevent a consensus among a group of agents by just varying the initial distribution of opinions in a case where the only dynamical mechanism is a pairwise bounded confidence interaction rule. For instance, systems with an initial distribution of opinions moderately polarized in two different opinion groups will find it much more difficult to arrive at a consensus, even if the individuals are willing to deal with very distant opinions. On the contrary, systems with an initial distribution of opinions moderately consensual will very easily find a globally shared opinion, regardless of how close-minded the agents are. In particular, we have shown that the transition from consensus to more than one opinion group can be moved in the range $\epsilon \in (0.135, 0.495)$ for the symmetric set of initial conditions we used, while the range is $\epsilon \in (0.245, 0.425)$ for the asymmetric initial distributions presented.

However, if the agents have some uniformly distributed preferred opinions and the ability to randomly go back to these preferences from time to time, then the possibility to initially prevent a consensus is totally removed and that of forcing it is substantially reduced. We have shown, both analytically and numerically, that the importance of the initial condition in determining the asymptotic state of the system is mainly replaced by the distribution of the noise or preferred opinions. Nevertheless, there is still a slight but noticeable impact of the initial condition upon the final or steady state, particularly for those initial distributions showing a moderate consensus. Thus, we observe in the noisy symmetric case that the transition from consensus to more than one opinion group can be a sharp transition at $\epsilon \approx 0.315$ or a smooth transition taking place in the range $\epsilon \in (0.235, 0.315)$ depending on the variance of the initial condition. In the noisy asymmetric case, we observe a change in the this transition in the much shorter range $\epsilon \in (0.295, 0.315)$.

Given the important differences observed in the noiseless model between the mass of the biggest cluster plots of the symmetric and the asymmetric sets of initial conditions, it is clear that the variance of the initial opinion distribution is not the correct or not the only parameter to take into account. As it can also be noticed in the related bifurcation diagrams, the variance is not enough to determine if the system will end up in a consensus or not for a given bound of confidence value. We notice that the symmetry of the initial condition does also play an important role. Therefore, we would probably need to take into account other parameters of the initial distribution as, for example, some higher moments.

Regarding the model with noise, the differences found in the bifurcation diagrams and the mass of the biggest cluster figures between both sets of initial conditions are not as important as in the original model. In fact, the bifurcation patterns observed in the mass of the biggest cluster figures show that the initial condition is irrelevant in determining the final configuration if the variance is strong enough, that is, if we sufficiently encourage the system to initially split into two main opinion groups at the extremes. Nevertheless, for lower values of the initial distribution variance, the final configuration of the system still depends on this initial condition. In particular, the variance is again unable to precisely determine the existence or not of a consensus, even though it gives much better prediction than in the noiseless case.

%

\appendix

\appendixpage

\numberwithin{equation}{section}

\section{Lyapunov function for the DW \textit{et al.} model}
\label{appendix}

In order to write a Lyapunov function for the DW \textit{et al.} process, we need to find a function $\mathcal{L}[\vec{x}(t)]$, where $\vec{x}(t)$ is a vector whose components are the agents opinions at time $t$, which satisfies
\begin{align}
 &\mathcal{L}[\vec{x}(t)] \ge 0 \quad \forall t,\label{a1}\\
 &\mathcal{L}[\vec{x}(t+1)] \le \mathcal{L}[\vec{x}(t)] \quad \forall t,\label{a2}
\end{align}
where \eqref{a1} simply means that it is a positive-definite function and \eqref{a2} that it cannot increase with time.

We will first write a positive-definite function and then prove that it is decreasing in time for the model interaction rules. Let us write a function which only depends on the distances between the agents opinions as
\begin{equation}
 \mathcal{L}[\vec{x}] = \sum_{i > j} (x_i - x_j)^2,
\end{equation}
and then focus on one interaction, i.e., on the changes occurred to the positions of only two agents in the opinion space, say agents $i_1$ and $j_1$. Thus, we may divide the Lyapunov function into two terms, one dependent and the other independent of $i_1$ and $j_1$, and let us call $A$ to the independent part for simplicity:
\begin{equation}
  \mathcal{L}[\vec{x}(t)] = A + \sum_{i\ne i_1,j_1}  (x_{i_1}(t) - x_i(t))^2 + \sum_{i\ne i_1,j_1}(x_{j_1}(t) - x_i(t))^2 + (x_{i_1}(t) - x_{j_1}(t))^2.
\end{equation}
Each sum contains $N-2$ terms, being $N$ the number of agents. Now we use the new opinions $x_{i_1}(t+1)$ and $x_{j_1}(t+1)$ that agents $i_1$ and $j_1$ hold after the interaction. It is important to notice that this interaction does only take place in case the opinions of the agents are nearer than the bound of confidence $\epsilon$. However, this does not affect our analysis, as we are only interested in effective interactions, those which actually take place. The Lyapunov function after the interaction is

\begin{equation}
  \mathcal{L}[\vec{x}(t)] = A + \sum_{i\ne i_1,j_1}  (x_{i_1}(t+1) - x_i(t))^2 + \sum_{i\ne i_1,j_1}(x_{j_1}(t+1) - x_i(t))^2 +(x_{i_1}(t+1) - x_{j_1}(t+1))^2.
\end{equation}
Replacing the new values $x_{i_1}(t+1)$ and $x_{j_1}(t+1)$ as given by application of the rule Eq.\eqref{modelDef} and subtracting, we get the variation of the Lyapunov as $\Delta \mathcal{L} =\mathcal{L}[\vec{x}(t+1)]-\mathcal{L}[\vec{x}(t)]$ which, after some algebra, reads:
\begin{equation}
 \Delta \mathcal{L} = -2\mu(1-\mu)N (x_{i_1} (t)- x_{j_1}(t))^2.
\end{equation}

Therefore, we see that the Lyapunov function $\cal L$ is strictly decreasing in time when any interaction takes place, and it stays constant when no interaction occurs.

{\bf Acknowledgements}

Financial support from  Ministerio de Ciencia e Innovaci\'on (Spain) and FEDER (EU) grant FIS2007-60327 is acknowledged.


\end{document}